\documentclass[aps,pra,reprint,longbibliography]{revtex4-1}
\usepackage{graphicx,color} 
\usepackage{amsmath, amssymb}
\usepackage{longtable}
\usepackage{natbib}
\usepackage{bm}

\begin{document}
\title{
Berry phase of adiabatic electronic configurations in fullerene anions
}
\author{Naoya Iwahara}
\email{naoya.iwahara@kuleuven.be}
\affiliation{Theory of Nanomaterials Group, University of Leuven, Celestijnenlaan 200F, B-3001 Leuven, Belgium}
\date{\today}

\begin{abstract}
The selection rule on vibronic angular momentum of $t_{1u}^n \otimes h_g$ Jahn-Teller problem ($n = $ 1-5) is reinvestigated. 
It is shown that among three adiabatic orbitals only two have nonzero Berry phase. 
Thus, the Berry phase of adiabatic electronic configurations depends on the spin multiplicity as well as the number of electrons. 
On this basis, the general relation between the Berry phase and the angular momentum is described. 
It allows us, in particular, to clarify the nature of vibronic states arising from high spin configurations. 
In comparison with the previous solution for the low-lying vibronic states for bimodal systems, the present solutions correctly fulfill all the symmetry requirement.
\end{abstract}

\maketitle

\section{Introduction}
\label{Introduction}
It is widely accepted that the Berry phase \cite{Berry1984} appears in various physical systems and characterizes their properties. 
In molecular physics, the Berry phase has been recognized as the sign change of an adiabatic orbital by the excursion along the path encircling the degenerate point in the nuclear configuration space \cite{Longuet-Higgins1958, Herzberg1963, OBrien1964, Mead1979, Mead1980}. 
Within the dynamic $E \otimes e$ Jahn-Teller (JT) problem, where the double degenerate $E$ electronic state couples to double degenerate $e$ normal mode, the phase change makes the angular momentum characterizing the vibronic states half-integer \cite{Longuet-Higgins1958, OBrien1964}, which has been experimentally addressed to evidence the Berry phase \cite{vonBusch1998, Keil2000}.
The relation between the Berry phase and the vibronic states was clarified by Ham \cite{Ham1987}.
He found in the $E \otimes e$ JT problem that the sequence of the irreducible representations of the vibronic states is modified by the presence of the Berry phase.
Since then the relation between the vibronic (also rovibronic) states and the Berry phase has been intensively studied 
\cite{Chancey1988, Aspel1992, Auerbach1994, OBrien1996, Moate1996, DeLosRios1996, Manolopoulos1999, Koizumi1999, Babikov2005, Garcia-Fernandez2006, Worner2006, Varandas2010, Pae2017, Requist2017}.

The majority of the works have addressed the simple cases where one electron or hole occupies degenerate orbitals, whereas Auerbach {\it et al}. \cite{Auerbach1994} studied the relation between the Berry phase and the population of the orbital shell as in the case of fullerene anions (C$_{60}^{n-}$, $n =$ 1-5). 
In their study of $t_{1u}^n \otimes h_g$ JT model, where triply degenerate $t_{1u}$ orbital couples to the five-fold degenerate $h_g$ vibrations, the following selection rule for the vibronic angular momentum $L$ was proposed \cite{Auerbach1994}: 
\begin{eqnarray}
 (-1)^{n + L} = 1.
\label{Eq:nL}
\end{eqnarray}
This relation indeed gives correct $L$ of low-spin ground vibronic states in the strong coupling limit. 
However, being applied to the description of the high-spin terms, it gives contradictory results. 
For example, the selection rule always predicts odd $L$ for $t_{1u}^1$ system and even $L$ for $t_{1u}^2$ system, whereas, spin triplet (high-spin) state of $t_{1u}^2$ system must have odd $L$ because the vibronic Hamiltonian is isomorphic to the one for $t_{1u}^1$. 
Although the isomorphism has been confirmed later \cite{OBrien1996, Liu2017}, the relation between the Berry phase and high-spin state has not been elucidated.

In this work, the selection rule for vibronic angular momentum in $t_{1u}^n \otimes h_g$ JT problem is revisited.
The symmetry properties of adiabatic orbitals are inspected, and the Berry phases of $t_{1u}^n$ electron configurations are established. 
On this basis, the selection rule on angular momentum (\ref{Eq:nL}) is generalized for 
both low- and high-spin adiabatic configurations. 
The present solution for the vibronic state of bimodal system is consistent with the symmetry requirement.

\section{Vibronic Hamiltonian}
In C$_{60}^{n-}$ anions with icosahedral symmetry, the triply degenerate $t_{1u}$ orbital couples to the five-fold degenerate $h_g$ normal mode \cite{Jahn1937}.
Although there are eight sets of JT active $h_g$ modes in C$_{60}$ molecule, the effective model including only one set of $h_g$ modes is considered. 
Moreover, the bielectronic interaction which induces the term splitting is neglected. 
The model vibronic Hamiltonian consists of the harmonic oscillator part $\hat{H}_0$ and the linear vibronic part $\hat{H}_\text{JT}$ \cite{Auerbach1994, OBrien1996, Chancey1997}:
\begin{eqnarray}
 \hat{H} = \hat{H}_0 + \hat{H}_\text{JT}.
\label{Eq:H}
\end{eqnarray}
The first contribution is the vibrational Hamiltonian for JT active modes 
\footnote{
In coordinate representation, $\hat{~}$ for operator is removed.
},
\begin{eqnarray}
 H_0 &=& \sum_{\gamma = \theta, \epsilon, \xi, \eta, \zeta} \frac{\hslash \omega}{2}
 \left(
  p_\gamma^2 + q_\gamma^2
 \right),
 \label{Eq:H0}
\end{eqnarray}
where, $q_\gamma$ and $p_\gamma$ are dimensionless normal coordinates and the conjugate momenta \cite{Auerbach1994}, $\omega$ is frequency,
and a real basis for the $h_g$ representation is used
\footnote{
$\theta, \epsilon, \xi, \eta, \zeta$ transform as $\frac{1}{\sqrt{6}}(2z^2-x^2-y^2)$, $\frac{1}{\sqrt{2}}(x^2-y^2)$, $\sqrt{2} yz$, $\sqrt{2} zx$, $\sqrt{2} xy$, respectively.
}.
The vibronic interaction is given by \cite{OBrien1969, Auerbach1994, OBrien1996, Chancey1997}:
\begin{eqnarray}
 \hat{H}_\text{JT} &=& 
 \sum_\sigma
 \hslash \omega g
 \left( \hat{c}^\dagger_{x \sigma}, \hat{c}^\dagger_{y \sigma}, \hat{c}^\dagger_{z \sigma} \right)
\nonumber\\
 &\times&
 \begin{pmatrix}
   \frac{1}{2} {q}_\theta - \frac{\sqrt{3}}{2} {q}_\epsilon & - \frac{\sqrt{3}}{2} {q}_\zeta & - \frac{\sqrt{3}}{2} {q}_\eta \\
   - \frac{\sqrt{3}}{2} {q}_\zeta & \frac{1}{2} {q}_\theta + \frac{\sqrt{3}}{2} {q}_\epsilon & - \frac{\sqrt{3}}{2} {q}_\xi \\
   - \frac{\sqrt{3}}{2} {q}_\eta  & - \frac{\sqrt{3}}{2} {q}_\xi & - {q}_\theta 
 \end{pmatrix}
 \begin{pmatrix}
  \hat{c}_{x \sigma}\\
  \hat{c}_{y \sigma}\\
  \hat{c}_{z \sigma} 
 \end{pmatrix},
\label{Eq:HJT}
\end{eqnarray}
where, $\hat{c}_{\gamma \sigma}^\dagger$ and $\hat{c}_{\gamma \sigma}$ are the electron creation and annihilation operators in $t_{1u}$ orbital $\lambda$ ($= x, y, z$) with $z$ component of electron spin $\sigma$ ($= \uparrow, \downarrow$), respectively, and $g$ is the dimensionless vibronic coupling parameter.  
Hereafter, for simplicity, the Hamiltonian is measured in units of $\hslash \omega$, and the phase factor of the normal mode is chosen so that $g > 0$.

The harmonic oscillator and the electronic Hamiltonian (orbital energy is set to zero) possess SO(5) and SO(3) symmetries, respectively, 
and the symmetry of the system is SO(5) $\otimes$ SO(3), which is reduced by the vibronic interaction to SO(3) symmetry \cite{OBrien1971}. 
The vibronic angular momentum $\hat{\bm{L}}$ is defined by the sum of vibrational part $\bm{L}_\text{vib}$ and electronic part $\hat{\bm{L}}_\text{el}$.
The Hamiltonian $\hat{H}$, the squared angular momentum $\hat{\bm{L}}^2$, and one of the components, e.g. $\hat{L}_z$, mutually commute.
Therefore, the vibronic state (eigen state of $\hat{H}$) is characterized by angular momentum $L$ and its $z$ component $M$.
In this work, the selection rule on the angular momentum $L$ is derived in the strong limit of the vibronic coupling, $g \gg 1$.

\section{Polar coordinates}
In the strong coupling limit, the adiabatic orbitals (eigen states of the matrix in $\hat{H}_\text{JT}$ (\ref{Eq:HJT})) give fundamental information of the low-energy electronic states.
For the diagonalization, it is convenient to introduce the polar coordinates \cite{OBrien1971, Auerbach1994, OBrien1996, Chancey1997}. 
Because of its importance, the parametrization of the $t_{1u}$ orbitals and the $h_g$ coordinates is described in detail below.

\begin{figure}[tb]
\begin{tabular}{lc}
 (a) 
\\
&
\includegraphics[width=5cm, bb=0 0 307 304]{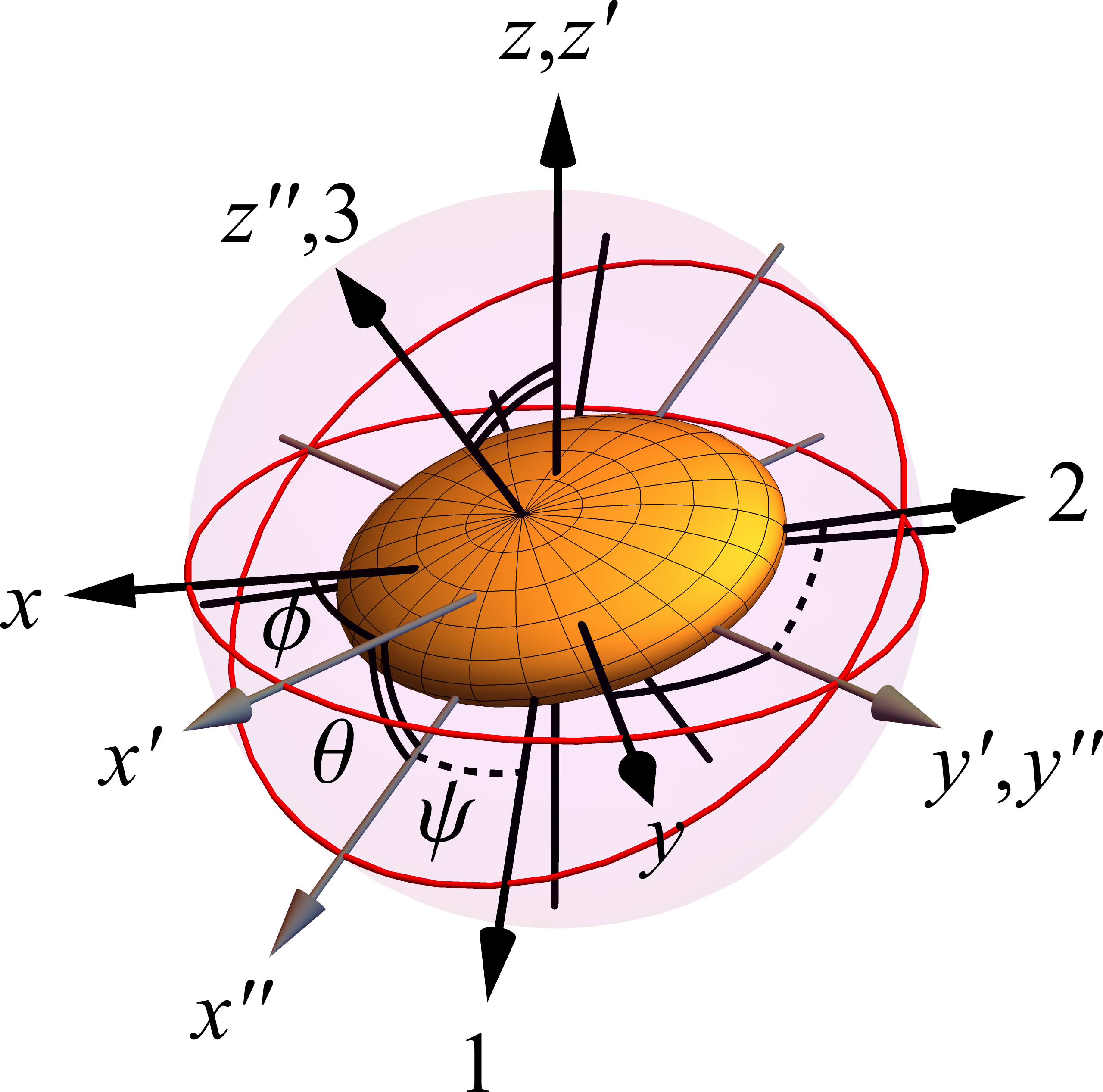}
\\
 (b) 
\\
&
\includegraphics[width=6cm]{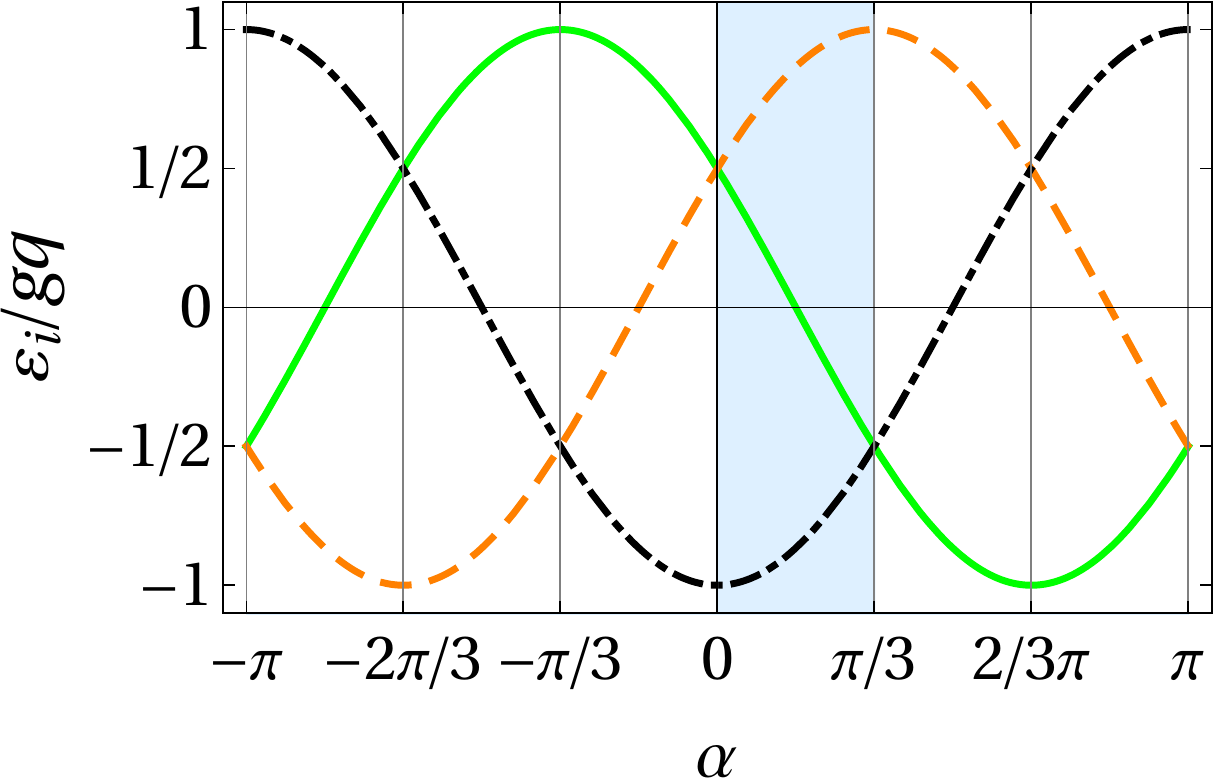}
\end{tabular}
\caption{
(Color online)
(a) Laboratory ($x,y,z$) and rotating ($1,2,3$) coordinate systems and Euler angles ($\phi, \theta, \psi$) describing the orientation of the spheroid (orange). 
The red circles show $x$-$y$ and $1$-$2$ planes. 
(b) Evolution of $\varepsilon_i$ (\ref{Eq:epsilon_spheroid}) with respect to angle $\alpha$. 
The green solid, orange dashed, and black dot dashed lines correspond to $\varepsilon_1$, $\varepsilon_2$, $\varepsilon_3$, respectively.
The blue shaded region appears six times in the plot.
}
\label{Fig:angle}
\end{figure}

We make use of the analogy between the derivation of adiabatic orbitals and that of the principal values of the inertia tensor of a rigid body, 
\begin{eqnarray}
 \sum_{\lambda, \lambda' = x, y, z} \omega_\lambda I_{\lambda \lambda'} \omega_{\lambda'} \rightarrow \sum_{i=1,2,3} I_i \omega_i^2,
\label{Eq:inertia}
\end{eqnarray}
where, $I_{\lambda \lambda'}$ is the inertia moment and $\omega_\lambda$ is $\lambda$ component of the angular velocity in the laboratory coordinate system $(x,y,z)$, 
and $I_i$ and $\omega_i$ ($i = 1, 2, 3$) are the principal moment of inertia and the angular velocity in the rotating coordinate system spanned by axes $i$ (Fig. \ref{Fig:angle}(a)) \cite{Landau1976}. 
Suppose both coordinate systems are right handed, they are related by the rotation defined by Euler angle $\Omega = (\phi, \theta, \psi)$ with the range of $0 \le \theta \le \pi$, $0 \le \phi, \psi < 2\pi$.
In the JT problem, Eq. (\ref{Eq:HJT}) and its diagonal form,
\begin{eqnarray}
 \hat{H}_\text{JT}
 &=& \sum_\sigma \sum_{i=1}^3 \varepsilon_i \hat{n}_{i\sigma},
\label{Eq:HJT_diag_0}
\end{eqnarray}
are, respectively, regarded as left and right hand sides of Eq. (\ref{Eq:inertia}). 
Thus, $x,y,z$ and $1,2,3$ axes correspond to the laboratory and rotating coordinates, 
the adiabatic orbital energy $\varepsilon_i$ to $I_i$, and the adiabatic orbitals,
\begin{eqnarray}
 \hat{a}_{i\sigma}^\dagger &=& \sum_{\lambda = x, y, z} \hat{c}_{\lambda \sigma}^\dagger {D}^{(1)}_{\lambda i}(\Omega),
\label{Eq:ad_orb}
\end{eqnarray}
to $\omega_i$.
Here, $D^{(1)}$ is Wigner $D$-function of rank 1 with the Condon-Shortley phase convention \cite{Varshalovich1988}, and $\hat{n}_{i\sigma} = \hat{a}^\dagger_{i\sigma} \hat{a}_{i\sigma}$. 
With the real basis of representations, Wigner $D$-functions are real. 

The parametrization of five dimensional coordinates in Ref. \cite{Bohr1952} is adapted to the JT problem. 
The obtained coordinates and the ranges are equivalent to those in Refs. \cite{OBrien1971, Auerbach1994, OBrien1996, Chancey1997}
\footnote{
The polar coordinates $(q, \alpha, \phi, \theta, \psi)$ in this work correspond to 
$(\beta, \gamma, \varphi, \theta, \psi)$ in Ref. \cite{Bohr1952}, 
$(q, \alpha, \phi, \theta, \frac{\beta}{2})$ in Ref. \cite{OBrien1971}, 
$(\sqrt{z^2+r^2}, \arctan\frac{r}{z}, \phi, \theta, \psi)$ in Ref. \cite{Auerbach1994}, 
$(q, \alpha, \phi, \theta, \gamma)$ in Ref. \cite{OBrien1996}, and 
$(Q, \alpha, \phi, \theta, \gamma)$ in Ref. \cite{Chancey1997}.
}.
In general, three $\varepsilon_i$'s differ from each other, whereas the trace of the matrix is invariant under rotation (i.e., $\sum_{i=1}^3\varepsilon_i = 0$), and thus, they are expressed by two parameters.
The deformation of the spheroid from sphere $(\varepsilon_1 = \varepsilon_2 = \varepsilon_3)$ is described by linear combination of the real parts of spherical harmonics $Y_{2m}$ \cite{Varshalovich1988} of rank 2 and component $m = 0, \pm 2$ in the rotating coordinate system:
\begin{eqnarray}
\begin{pmatrix}
 \varepsilon_1\\ 
 \varepsilon_2\\ 
 \varepsilon_3 
\end{pmatrix}
&=& 
 a'_\theta \bm{e}'_\theta + a'_\epsilon \bm{e}'_\epsilon
 =
 g q
 \begin{pmatrix}
  -\cos \left(\alpha - \frac{2\pi}{3} \right)\\
  -\cos \left(\alpha + \frac{2\pi}{3} \right)\\
  -\cos \alpha
 \end{pmatrix}.
\label{Eq:epsilon_spheroid}
\end{eqnarray}
Here, $\bm{e}'_\theta = \frac{1}{\sqrt{6}}(-1, -1, 2)^T$ and $\bm{e}'_\epsilon = \frac{1}{\sqrt{2}}(1, -1, 0)^T$ are the $\theta$ ($Y_{20}$) and $\epsilon$ ($\frac{1}{\sqrt{2}}[Y_{2,-2}+Y_{22}]$) types of deformations, respectively, 
the coordinates $a'_\kappa$ ($\kappa = \theta, \epsilon$) are parametrized as $a'_\theta = -\frac{\sqrt{6}}{2} g q \cos \alpha$, and  $a'_\epsilon = -\frac{\sqrt{6}}{2} g q \sin \alpha$.
The coefficient ($-\frac{\sqrt{6}}{2}$) and $g$ are introduced by comparing the forms of the Hamiltonians (\ref{Eq:HJT}) and (\ref{Eq:HJT_diag_0}) in the special case where two coordinate systems coincide ($x = 1, y = 2, z = 3$).
In this case, $q$ corresponds to the magnitude of the deformation and $\alpha$ determines the direction in the $q_\theta$-$q_\epsilon$ plane ($0 \le q$ and $-\pi \le \alpha < \pi$).
The other coordinates ($\kappa = \xi, \eta, \zeta$) do not contribute because they only modify the angles between the principal axes. 
Similarly, the shape of the rigid body is written as $\sum_{\gamma} q_\gamma \bm{e}_\gamma$ in the laboratory coordinate system,
where, the polarization vectors $\bm{e}_\gamma$'s also express the deformation of rank 2.
The polarization vectors $\bm{e}'_\kappa$ are obtained by rotation of the polarization vectors $\bm{e}_\gamma$ in the laboratory coordinate system,
$\bm{e}'_\kappa = \sum_\gamma \bm{e}_\gamma {D}^{(2)}_{\gamma \kappa}(\Omega)$,
and hence, the normal coordinates $q'_\kappa$ ($= \frac{2}{\sqrt{6}g} a'_\kappa$) and $q_\gamma$ are related by the inverse rotation:
\begin{eqnarray}
 q_\gamma &=& \sum_{\kappa = \theta, \epsilon} {D}^{(2)}_{\gamma \kappa}(\Omega) q'_\kappa.
\label{Eq:q}
\end{eqnarray}

It results that the five-dimensional $h_g$ coordinates are described by the shape ($q, \alpha$) and the orientation ($\Omega$) of the rigid body (Fig. \ref{Fig:angle}(a)).
However, the Cartesian $q_\gamma$ and polar $(q, \alpha, \Omega)$ coordinates are not in a one-to-one correspondence because the latter have extra degrees of freedom to label the principal axes $(1,2,3)$ and to fix their directions (Fig. \ref{Fig:angle}(a) is an example).
The degrees of freedom amount to as many as 48 ways ($\cong O_h$) given the numbers of combinations are $3!=6$ for the former and $2^3=8$ for the latter. 
By choosing the labels and the directions, the ranges of the polar coordinates are restricted.
Different labeling of the axes for certain set of orbital energies (\ref{Eq:epsilon_spheroid}) is achieved by shifting $\alpha$.
Figure \ref{Fig:angle}(b) shows that there are six physically equivalent regions of $\alpha$ with different order of $\varepsilon_i$'s, and one of the equivalent domains has to be chosen.
Now, there remain only 8 ways of arbitrariness regarding the directions of axes ($\cong D_{2h}$). 
By taking right handed coordinate system, which discards the inversion of the coordinate systems, the degrees of freedom are further decreased to 4 ($\cong D_2$).
Choosing one set of directions is equivalent to the restriction of the domain of Euler angles. 
The restriction can be done by using the generators of $D_2$ group, for example, the $C_2$ (or $\pi$) rotation around principal axis 1 ($R_1$) and the $C_2$ rotation around axis 3 ($R_3$). 
Under $R_1$, the Euler angles change as (see e.g. Ref. \cite{Bohr1952} and Sec. 4.4 in Ref. \cite{Varshalovich1988})
\begin{eqnarray}
 (\phi, \theta, \psi) \rightarrow (\phi + \pi, \pi - \theta, -\psi).
\label{Eq:R1}
\end{eqnarray}
Since both angles express physically the same orientations of the rigid body ($\varepsilon_i$'s)
with different directions of the principal axes 2 and 3, their domains should not overlap.
Consequently, the range of $\theta$ is narrowed as $0 \le \theta \le \frac{\pi}{2}$.
By the second rotation $R_3$, the Euler angles becomes 
\begin{eqnarray}
 (\phi, \theta, \psi) \rightarrow (\phi, \theta, \psi + \pi),
\label{Eq:R3}
\end{eqnarray}
and the range of $\psi$ decreases to $0 \le \psi < \pi$.
As the result, the Cartesian coordinates $q_\gamma$ and the polar coordinates are in one-to-one correspondence when
$0 \le q$, $0 \le \alpha \le \frac{\pi}{3}$ or in the other equivalent domain, 
$0 \le \phi < 2\pi$, $0 \le \theta \le \frac{\pi}{2}$, $0 \le \psi < \pi$.

Figure \ref{Fig:angle}(a) uniquely defines the adiabatic orbitals and normal coordinates within the obtained domain:
the directions of the principal axes of the rigid body correspond to the three adiabatic orbitals, and the shape and the orientation of the rigid body show the radial coordinates $(q,\alpha)$ and Euler angles $(\phi, \theta, \psi)$ of the $h_g$ normal coordinate, respectively.
The domain of coordinates allows any shape and orientation of the rigid body such that the ellipsoid retains the order of the principal moment of inertia and the directions of the principal axes do not coincide with the other ones. 

With the polar coordinates defined above, the kinetic energy term splits into radial vibrational part and pseudorotational part \cite{Bohr1952}, 
the potential energy term of $H_0$ becomes $q^2/2$, and the vibronic term (\ref{Eq:HJT_diag_0}) contains only radial coordinates.
The pseudorotational part is expressed as 
\begin{eqnarray}
 H_\text{rot} = \sum_{i=1}^3 \frac{L_i^2}{2q^2 I_i}, 
\label{Eq:Hrot}
\end{eqnarray}
where, ${I}_i$ is the principal moment of inertia: 
\begin{eqnarray}
 {I}_i &=& 4 \sin^2\left(\alpha - i \frac{2\pi}{3}\right),
\label{Eq:I}
\end{eqnarray}
and $L_i$ are the angular momentum of the rigid body in the rotating coordinate system \cite{Casimir1931}.
One should note that ${I}_i$ in Eq. (\ref{Eq:I}) and $I_i$ in Eq. (\ref{Eq:inertia}) used for the analogy with $\varepsilon_i$ have nothing in common. 
$L_i$ corresponds to the vibrational angular momentum in the rotating coordinate system, and also to the vibronic angular momentum $L$ within adiabatic approximation
\footnote{
In the rotating coordinate system (\ref{Eq:ad_orb}), the electronic angular momentum $\hat{L}_\text{el}$ does not have diagonal elements, and its expectation value in an adiabatic electronic state is zero. 
The expectation values of $(\hat{\bm{L}}_\text{vib})^2$ and $(\hat{\bm{L}}_\text{el})^2$ in adiabatic electronic state cancel each other. 
Therefore, the vibronic angular momentum acting on the nuclear part of the vibronic state is expressed by $L_i$.
}.

\begin{table*}[tb]
\caption{
The JT distortion, static JT energy, typical adiabatic states, Berry phases under rotation $R_1$ (\ref{Eq:R1}) and $R_3$ (\ref{Eq:R3}), and moments of inertia $I_i$. 
}
\label{Table:berry}
\begin{ruledtabular}
\begin{tabular}{ccccccccc}
 $n$ & $S$ & $(q, \alpha)$ & $\varepsilon_i$ & $E_\text{JT}$ & $|\Phi^{(n)}\rangle$ & \multicolumn{2}{c}{Berry phase} & ${I}_i$\\
 & & & & & & $R_1$ & $R_3$ \\
\hline
 1 & $\frac{1}{2}$ & $(g, 0)$        & $\varepsilon_3 < \varepsilon_1 = \varepsilon_2$ & $-\frac{1}{2}g^2$ & $\hat{a}_{3\sigma}^\dagger|0\rangle$ 
   & $-1$ & 1 & (3,3,0) \\
 2 & 0   & $(2g, 0)$       & $\varepsilon_3 < \varepsilon_1 = \varepsilon_2$ & $-2g^2$ & $\hat{a}_{3\uparrow}^\dagger \hat{a}_{3\downarrow}^\dagger |0\rangle$ 
   & 1 & 1 & (3,3,0) \\
   & 1   & $(g, \pi)$      & $\varepsilon_1 = \varepsilon_2 < \varepsilon_3$ & $-\frac{1}{2}g^2$ & $\hat{a}_{1\sigma}^\dagger \hat{a}_{2\sigma}^\dagger |0\rangle$ 
   & $-1$ & 1 & (3,3,0) \\
 3 & $\frac{1}{2}$ & $(\sqrt{3}g, \frac{\pi}{2})$ & $\varepsilon_1 < \varepsilon_3 < \varepsilon_2$ & $-\frac{3}{2}g^2$ & $\hat{a}_{1\uparrow}^\dagger \hat{a}_{1\downarrow} ^\dagger \hat{a}_{3\sigma}^\dagger |0\rangle$ 
   & $-1$ & 1 & (1,1,4)\\
   & $\frac{3}{2}$ & (0, -)          & $\varepsilon_i = 0$ & 0 & $\hat{a}_{1\sigma}^\dagger \hat{a}_{2\sigma}^\dagger \hat{a}_{3\sigma}^\dagger |0\rangle$ 
   & 1 & 1 & - \\
 4 & 0   & $(2g, \pi)$     & $\varepsilon_1 = \varepsilon_2 < \varepsilon_3$ & $-2g^2$ & $\hat{a}_{1\uparrow}^\dagger \hat{a}_{1\downarrow}^\dagger \hat{a}_{2\uparrow} ^\dagger \hat{a}_{2\downarrow}^\dagger |0\rangle$ 
   & 1 & 1 & (3,3,0) \\
   & 1   & $(g, 0)$        & $\varepsilon_3 < \varepsilon_1 = \varepsilon_2$ & $-\frac{1}{2}g^2$ & $\hat{a}_{1\sigma}^\dagger \hat{a}_{2\sigma}^\dagger \hat{a}_{3\uparrow} ^\dagger \hat{a}_{3\downarrow}^\dagger |0\rangle$ 
   & $-1$ & 1 & (3,3,0)\\
 5 & $\frac{1}{2}$ & $(g, \pi)$      & $\varepsilon_1 = \varepsilon_2 < \varepsilon_3$ & $-\frac{1}{2}g^2$ & $\hat{a}_{1\uparrow}^\dagger \hat{a}_{1\downarrow}^\dagger \hat{a}_{2\uparrow}^\dagger \hat{a}_{2\downarrow} ^\dagger \hat{a}_{3\sigma}^\dagger |0\rangle$ 
   & $-1$ & 1 & (3,3,0) \\
\end{tabular}
\end{ruledtabular}
\end{table*}

\begin{figure*}[tb]
\includegraphics[width=4.5cm, angle=-90]{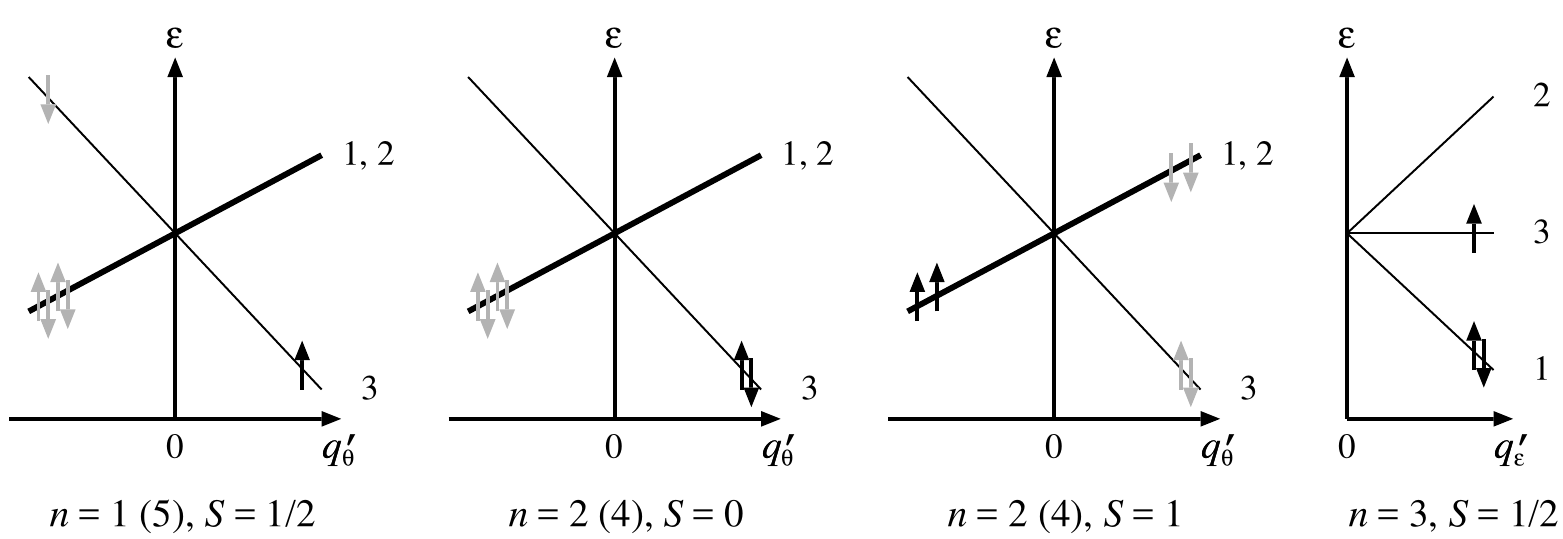}
\caption{
Electron configuration for different $n$ and $S$. 
The thin and thick slopes indicate the nondegenerate and doubly degenerate adiabatic orbitals, respectively,
$q'_\theta$ and $q'_\epsilon$ are the normal coordinate in the rotating coordinate, 
the black arrows are electrons for $n \le 3$ and gray arrows indicate those for $n \ge 4$.
}
\label{Fig:Phi}
\end{figure*}

\section{Berry phase}
\subsection{Adiabatic orbitals}
The phase change of each adiabatic orbital (\ref{Eq:ad_orb}) is examined under the adiabatic process around the circuit in the nuclear coordinate space. 
Pictorially, this can be done by observing the evolution of axes $i$ under the adiabatic transportation such that the initial and final orientations of the spheroid coincide (Fig. \ref{Fig:angle}(a)). 
For any loops described inside the domain of Euler angles, the axes simply return to their original positions.
On the other hand, for any paths connecting two sets of Euler angles related by $C_2$ rotation with respect to axis $i$, 
two axes except for $i$ are reversed at the end point with respect to those of the starting point. 
The reversal of the direction of axis $i$ means the sign change of adiabatic orbital $i$ 
\footnote{
This is easily understood in the case of $E \otimes e$ JT model, where 
the adiabatic orbitals with energies $g\rho$ and $-g\rho$ are \unexpanded{$|+\rangle = \cos \frac{\phi}{2} |\theta\rangle - \sin \frac{\phi}{2} |\epsilon\rangle$}
and \unexpanded{$|-\rangle = \sin \frac{\phi}{2} |\theta\rangle + \cos \frac{\phi}{2} |\epsilon\rangle$}, respectively, 
and the coordinates are \unexpanded{$q_\theta = q \cos \phi, q_\epsilon = q \sin \phi$}.
\unexpanded{$|\theta\rangle, |\epsilon\rangle$} and \unexpanded{$|\pm \rangle$} correspond to the unit vectors along the axes of the laboratory and the rotating coordinates, respectively. 
Under the $2\pi$ rotation, the coordinates returns to the original position, whereas the directions of the principal axes are opposite to the original ones. 
The latter indicates the sign change of adiabatic orbitals. 
}.
The second types of path is also closed in nuclear configuration space because $C_2$ is symmetry operation of the spheroid. 

The phase change can be demonstrated more directly by using the explicit form of adiabatic orbitals,
\begin{widetext}
\begin{eqnarray}
\hat{a}_{1\sigma}^\dagger
 &=& 
  (\cos \psi \cos \theta \cos \phi - \sin \psi \sin \phi) \hat{c}_{x\sigma}^\dagger
+ (\cos \psi \cos \theta \sin \phi + \sin \psi \cos \phi) \hat{c}_{y\sigma}^\dagger
-  \cos \psi \sin \theta                                  \hat{c}_{z\sigma}^\dagger,
\nonumber\\
\hat{a}_{2\sigma}^\dagger 
 &=&
- (\sin \psi \cos \theta \cos \phi + \cos \psi \sin \phi) \hat{c}_{x\sigma}^\dagger
- (\sin \psi \cos \theta \sin \phi - \cos \psi \cos \phi) \hat{c}_{y\sigma}^\dagger
+  \sin \psi \sin \theta                                  \hat{c}_{z\sigma}^\dagger,
\nonumber\\
\hat{a}_{3\sigma}^\dagger 
 &=&
  \sin \theta \cos \phi  \hat{c}_{x\sigma}^\dagger
+ \sin \theta \sin \phi  \hat{c}_{y\sigma}^\dagger
+ \cos \theta            \hat{c}_{z\sigma}^\dagger.
\label{Eq:a_ad_2}
\end{eqnarray}
\end{widetext}
Under the first type of loops the resultant phase changes are null. 
For the second kind of loops, the sign changes result: 
with $R_1$ (\ref{Eq:R1}), the adiabatic orbitals transform as 
\begin{eqnarray}
 \left(\hat{a}_1^\dagger, \hat{a}_2^\dagger, \hat{a}_3^\dagger\right) 
 \rightarrow 
 \left(\hat{a}_1^\dagger, -\hat{a}_2^\dagger, -\hat{a}_3^\dagger\right),
\label{Eq:a_R1}
\end{eqnarray}
and with $R_3$ (\ref{Eq:R3}), 
\begin{eqnarray}
 \left(\hat{a}_1^\dagger, \hat{a}_2^\dagger, \hat{a}_3^\dagger\right) 
 \rightarrow 
 \left(-\hat{a}_1^\dagger, -\hat{a}_2^\dagger, \hat{a}_3^\dagger\right).
\label{Eq:a_R3}
\end{eqnarray}
Indeed, for $C_2$ rotation around axis $i$, all the adiabatic orbitals except for $i$ change their signs. 
This result is consistent with the fact that ${D}^{(1)}$ is an element of SO(3) group, i.e. $\det D^{(1)} = 1$ \cite{Manolopoulos1999} (see also Ref. \cite{Varandas2010}).

At a first glance, the difference in Berry phase of adiabatic orbitals seems to be contradictory to the physical equivalence of adiabatic orbitals. 
The equivalence can be seen in the symmetric situation under all three rotations in $D_2$:
by the third rotation $R_2 = R_1 \cdot R_3$, the adiabatic orbitals change as 
\begin{eqnarray}
 \left(\hat{a}_1^\dagger, \hat{a}_2^\dagger, \hat{a}_3^\dagger\right) 
 \rightarrow 
 \left(-\hat{a}_1^\dagger, \hat{a}_2^\dagger, -\hat{a}_3^\dagger\right),
\label{Eq:a_R2}
\end{eqnarray}
and now the sign of orbital 2 remains the same and the rests change.

\subsection{Adiabatic electron configurations}
The Berry phase of adiabatic electron configuration,
\begin{eqnarray}
 |\Phi^{(n)}(\Omega)\rangle &=& 
 \hat{a}_{i_1\sigma_1}^\dagger 
 \hat{a}_{i_2\sigma_2}^\dagger 
 \cdots 
 \hat{a}_{i_n\sigma_n}^\dagger 
 |0\rangle,
\label{Eq:Phi_ad}
\end{eqnarray}
for the loops including $R_i$ rotation is the product of those of occupied orbitals, 
\begin{eqnarray}
 |\Phi^{(n)}(\Omega)\rangle \rightarrow (-1)^{n-n_i} |\Phi^{(n)}(\Omega)\rangle,
\label{Eq:Phi_berry}
\end{eqnarray}
where $|0\rangle$ is the vacuum state of the $t_{1u}$ shell, and $n_i$ $(= 0,1,2)$ is the occupation number of electrons in adiabatic orbital $i$.
Thus, the main task here is the calculations of the lowest electron configurations for different $n$ and spin multiplicity. 
To this end, the potential terms of $\hat{H}$, 
\begin{eqnarray}
 \hat{U} &=& \frac{q^2}{2} - \sum_{\sigma} g q \left[
    \cos \left(\alpha - \frac{2\pi}{3} \right) \hat{n}_{1\sigma} 
    \right.
\nonumber\\
&+& 
    \left.
    \cos \left(\alpha + \frac{2\pi}{3} \right) \hat{n}_{2\sigma} 
 +  \cos \alpha \hat{n}_{3\sigma} \right],
\label{Eq:U}
\end{eqnarray}
are minimized with respect to $q$, $\alpha$, and electron occupations under constraint on $n$ and spin $S$ \cite{Auerbach1994},
where $\hat{U}$ is obtained using, Eqs. (\ref{Eq:HJT_diag_0}) and (\ref{Eq:epsilon_spheroid}).
The angle $\alpha$ is chosen so that $I_i$ (\ref{Eq:I}) becomes symmetric with respect to the third principal axis, ${I}_1 = {I}_2 \ne {I}_3$. 
The results are listed in Table \ref{Table:berry} and the electron configurations are shown in Fig. \ref{Fig:Phi}.

Under $R_3$, the sign of adiabatic electronic state does not change for any $n$ and $S$. 
Under $R_1$, orbital 1 does not change the sign, whereas the orbital is either empty or doubly filled as long as the low-spin adiabatic states are considered (Fig. \ref{Fig:Phi}). 
Therefore, the sign changes of low-spin states coincide with the electronic part, $(-1)^n$, of Eq. (\ref{Eq:nL}).
On the other hand, orbital 1 is occupied by one electron in high-spin states, $^3T_{1g}$ of $t_{1u}^{2}$ ($t_{1u}^4$) and $^4A_u$ of $t_{1u}^{3}$, and thus, the phase changes of these states are $-1$ and 1, respectively. 
The results differ from the prediction of Eq. (\ref{Eq:nL}).
The absence of the sign change for $^4A_u$ configuration is consistent with the nondegeneracy of the orbital part (no conical intersection).

\section{Selection rule on vibronic angular momentum}
The selection rule on vibronic angular momentum characterizing the low-energy vibronic states is further generalized so that it is applicable to both low- and high-spin cases.
As shown in Table \ref{Table:berry}, there are two types of moments of inertia: the oblate type (${I}_1 = {I}_2 > {I}_3$) and the prolate type (${I}_1 = {I}_2 < {I}_3$), which are, respectively, called unimodal and bimodal \cite{Auerbach1994}.
These cases are treated separately.

\subsection{Unimodal case}
Since one of the moments of inertia is zero (${I}_3 = 0$) as in diatomic molecule, the pseudorotational Hamiltonian $H_\text{rot}$ describes one-dimensional vibration and rotation of the system with two angles \cite{Landau1977} (See for detailed derivation Refs. \cite{Auerbach1994, OBrien1996}).
In this case, the vibronic state is given by the product of the adiabatic electronic, vibrational, and pseudorotational parts
\footnote{
The last part is in general $D_{MK}^{(L)}(\phi, \theta, 0)$ \cite{Landau1977}. $K = 0$ is chosen to obtain low energy states (see Eqs. (12) and (24) in Ref. \cite{OBrien1996}). 
$D_{M0}^{(L)}(\phi, \theta, 0)$ corresponds to spherical harmonics $Y_{LM}(\theta, \phi)$.
}:
\begin{eqnarray}
 |\Psi^{(n)}(q, \alpha, \Omega)\rangle &=& |\Phi^{(n)}(\Omega)\rangle \chi(q, \alpha, \psi) {D}_{M0}^{(L)}(\phi, \theta, 0).
\label{Eq:Psi_prolate}
\end{eqnarray}
The vibronic states have to be single valued under $R_1$ and $R_3$ (and thus, $R_2$). 
The sign changes of the adiabatic electronic configurations are $(-1)^{n-n_i}$, Eq. (\ref{Eq:Phi_berry}).
The pseudorotational part changes by $(-1)^L$ for $R_1$ and remains the same for $R_3$
\footnote{
Under $R_1$, ${D}^{(L)}_{MK}(\Omega) \rightarrow (-1)^L {D}^{(L)}_{M,-K}(\Omega)$,
and under $R_3$, ${D}^{(L)}_{MK}(\Omega) \rightarrow (-1)^K {D}^{(L)}_{MK}(\Omega)$ \cite{Varshalovich1988}.
}.
The radial vibrational part $\chi$ should be invariant under any changes of $\psi$ by $R_i$.
Therefore, the relation between the vibronic angular momentum $L$ and the Berry phase of adiabatic electronic state is given by 
\begin{eqnarray}
 (-1)^{n - n_1 + L} = 1.
\label{Eq:selection_1}
\end{eqnarray} 
In this formula, $n_1$ is equivalent to $n_2$ for the present choice of $\alpha$ which fulfills $I_1 = I_2 > I_3$. 
The latter corresponds to using $R_2$ rotation.
For other choice of $\alpha$, $n_1$ in Eq. (\ref{Eq:selection_1}) should be replaced by $n_i$ where $i$ is not the main symmetry axis of principal moment of inertia, $I_i = I_j \ne I_k$ ($i,j,k = 1,2,3$).

For the low-spin states with odd (even) number of electrons, the angular momentum $L$ is odd (even). 
This is in line with the previous results \cite{Auerbach1994}.
Contrary, the Berry phase of $^3T_{1g}$ term ($n = 2, 4$) is $-1$ as in the case of $^2T_{1u}$ term ($n = 1, 5$), thus the angular momentum $L$ ($L = 1$ for the ground state) is also odd. 
Since the Berry phase of the high-spin terms is adequately described, the angular momentum $L$ is consistent with the fact that the JT Hamiltonian for $^3T_{1g}$ term of $t_{1u}^2$ ($t_{1u}^4$) has the same form as that for $t_{1u}^1$. 
Although $^4A_u$ state of $t_{1u}^3$ is neither unimodal nor bimodal, by the similar discussion $L$ must be even, which is nothing but a vibrational angular momentum \cite{Bohr1952}.
The fingerprint of the vibronic angular momentum of the high-spin vibronic states could be observed by spectroscopic techniques and also in thermodynamic quantities.
An example of the latter is spin-gap of C$_{60}$ anions in which the large entropy due to the degeneracy of the low-lying vibronic states plays crucial role \cite{Liu2017}.

\subsection{Bimodal case}
Since all the moments of inertia are nonzero, the pseudorotational part of $H_0$ reduces to the Hamiltonian for a symmetric top with $\alpha = \frac{\pi}{2}$ 
\cite{Landau1977, Auerbach1994, OBrien1996}.
Similarly to the unimodal case, the vibronic state would be written as $|\Phi^{(3)}(\Omega)\rangle \chi_K(q, \alpha) D^{(L)}_{MK}(\Omega)$. 
This gives rise to $2(2L+1)$-fold degeneracy ($2L+1$ is from $M$ and 2 from $\pm K$ for $K \ne 0$), which is two times larger than expected, as pointed out by O'Brien \cite{OBrien1996}. 
This issue may be solved by taking into account quantum correction to the pseudorotational Hamiltonian due to the deviation of $\alpha$ from classical value $\frac{\pi}{2}$,
\begin{eqnarray}
 H_\text{rot} &=& \frac{1}{2q^2} 
 \left( \bm{L}^2 - \frac{3L_3^2}{4} \right)
 + \frac{\sqrt{3}}{q^2} \Delta \alpha  
 \left(L_1^2 - L_2^2\right).
\label{Eq:Hrot_bimodal}
\end{eqnarray}
where, $\bm{L}^2 = \sum_{i=1}^3 L_i^2$ and $\Delta \alpha = \alpha - \frac{\pi}{2}$.
Because of the second term, the system becomes asymmetric top being coupled to the radial part, and hence the vibronic states may be described by the superposition of pseudorotational states of symmetric top (see for asymmetric top Refs. \cite{Casimir1931, Bohr1952, Landau1977}).
Considering the transformation properties of the vibronic states under $C_2$ rotations, we obtain the selection rule as well as the form of the vibronic state (see Appendix \ref{A:Psi3}):
\begin{widetext}
\begin{eqnarray}
 |\Psi^{(3)}(q,\alpha,\Omega)\rangle &=& |\Phi^{(n)}(\Omega)\rangle \chi_{K}(q, \alpha) 
 \times 
 \begin{cases}
  \left(
  D_{MK}^{(L)}(\Omega) + D_{M,-K}^{(L)}(\Omega)
  \right)
  & L : \text{odd}, K = 0, 2, 4, ..., L-1 \\
  \left(
  D_{MK}^{(L)}(\Omega) - D_{M,-K}^{(L)}(\Omega)
  \right)
  & L: \text{even}, K = 2, 4, ..., L
 \end{cases}.
\label{Eq:Psi_oblate}
\end{eqnarray}
\end{widetext}
The selection rule agrees with Ref. \cite{OBrien1996}, however, the pseudorotational state differs from the previous one. 
The previous pseudorotational state shows sign change, while it is also suffered from unnecessary change of its form under symmetry operation. 
This issue is fixed in our solution.

\section{Conclusion}
The selection rule on vibronic angular momentum of $t_{1u}^n \otimes h_g$ Jahn-Teller problem is thoroughly investigated.
The Berry phases of adiabatic orbitals are analyzed, and it was confirmed that only two among three orbitals can have nonzero Berry phases. 
The Berry phase of adiabatic electronic configuration depends not only on the number of electrons $n$ but also on spin multiplicity.
On this basis, the relation between the Berry phase and vibronic angular momentum $L$ is generalized.
Particularly, the contradiction between the vibronic state of high-spin system and the selection rule is solved. 
In addition, the present vibronic state of bimodal system fulfills the symmetry requirement, which is not fully satisfied in the previous one. 
Present result enables the comprehensive understanding of the relation between the nature of the vibronic states of $t_{1u}^n \otimes h_g$ Jahn-Teller system.
Moreover, the dependence of Berry phase on configuration will play an important role in the other multielectronic Jahn-Teller systems such as $d$ metal ions in cubic site.

\section*{Acknowledgment}
N.I. is grateful to Liviu F. Chibotaru for stimulating discussions on this problem. 
N.I. is supported by Japan Society for the Promotion of Science (JSPS) Overseas Research Fellowship.

\appendix
\section{Calculation of Eq. (\ref{Eq:Psi_oblate})}
\label{A:Psi3}
The vibronic state of asymmetric top may be written as 
\begin{eqnarray}
 |\Psi^{(n)}\rangle &=& |\Phi^{(3)}(\Omega)\rangle \sum_{N = -L}^L \chi_N(q, \alpha) D_{MN}^{(L)}(\Omega). 
\end{eqnarray}
Since the vibronic state is single-valued in the space of nuclear coordinates, under the projection operator onto totally symmetric representation, 
\begin{eqnarray}
 \hat{P}_A &=& \frac{1}{|D_2|} \sum_{G \in D_2} \hat{G},
\end{eqnarray}
it remains the same, where, $|D_2|$ is the number of elements in the group. 
The projected state is 
\begin{eqnarray}
\hat{P}_A|\Psi^{(n)}\rangle 
 &=& 
 \frac{1}{4} |\Phi^{(3)}(\Omega)\rangle \sum_{N = -L}^L \chi_N(q, \alpha) \left( 1+(-1)^N \right)
\nonumber\\
 &\times&
 \left(
  D_{MN}^{(L)}(\Omega) - (-1)^L D_{M,-N}^{(L)}(\Omega)
 \right),
\end{eqnarray}
implying that $N$ has to be even. Expanding the sum over $N$ partly, 
\begin{eqnarray}
\hat{P}_A|\Psi^{(n)}\rangle 
 &=& 
 \frac{1}{2} |\Phi^{(3)}(\Omega)\rangle \sideset{}{'}\sum_{K} 
\nonumber\\
 &\times&
 \left[
  \chi_K^\text{S}
  \left( 1 - (-1)^L \right)
  \left(
  D_{MK}^{(L)}(\Omega) + D_{M,-K}^{(L)}(\Omega)
  \right)
 \right.
\nonumber\\
 &+&
 \left.
  \chi_K^\text{A}
  \left( 1 + (-1)^L \right)
  \left(
  D_{MK}^{(L)}(\Omega) - D_{M,-K}^{(L)}(\Omega)
  \right)
 \right],
\nonumber\\
\end{eqnarray}
where, $K$ is even and fulfills $0 \le K \le L$, and $\chi^\text{S}_K$ and $\chi^\text{A}_K$ are symmetric and antisymmetric part of $\chi_K$ under inversion of $K$.
For odd and even $L$, the first and second terms in the curly bracket become zero, respectively, 
and for $K = 0$, the second term is zero.
The dominant part for each vibronic state is shown in Eq. (\ref{Eq:Psi_oblate}).
In the expression, superscript of $\chi_K$ is omitted.



%

\end{document}